\documentstyle[12pt]{article} %available in the standard LaTex package.

\normalbaselines

\begin{document}
\bibliographystyle{unsrt}

\vbox {\vspace{6mm}} %Leave space at the top on the first page.

\begin{center}{
{\large \bf
DECOHERENCE AND DISSIPATION FOR A QUANTUM SYSTEM COUPLED TO A LOCAL ENVIRONMENT
}{\footnote{To appear in the Proceedings of the Third International Workshop on
Squeezed States and Uncertainty Relations, University of Maryland Baltimore
County}} \\[7mm]
Michael R. Gallis\\
{\it Department of Physics, Penn State
University/Schuylkill Campus\\
Schuylkill Haven, Pennsylvania 17972\\
Internet:mrg3@psuvm.psu.edu}
}\end{center}

\vspace{2mm}

\begin{abstract}
Decoherence and dissipation in quantum systems has been studied extensively
 in the context of Quantum Brownian Motion.  Effective decoherence in coarse
 grained quantum systems has been a central issue in recent efforts by Zurek
 and by Hartle and Gell-Mann to address the Quantum Measurement Problem.
%Typically, the models considered consist of a particle coupled to an
%oscillator bath, where the coupling term in the potential is linear in both
%the system and bath coordinates.
Although these models can yield very general
 classical phenomenology, they are incapable of reproducing relevant
 characteristics expected of a local environment on a quantum system,
 such as the characteristic dependence of decoherence on environment
 spatial correlations.  I discuss the characteristics of Quantum
 Brownian Motion in a local environment by examining aspects of first
 principle calculations and by the construction of phenomenological models.
  Effective quantum Langevin equations and master equations are presented
 in a variety of representations.  Comparisons are made with standard
 results such as the Caldeira-Leggett master equation.
\end{abstract}

\section{Introduction and Motivation}

  Decoherence via coarse
graining has been studied in the context of quantum measurement theory by
 Zurek\cite{zurek} and by Hartle
and Gell-Mann\cite{hng} as a mechanism which leads to the emergence of
classical
 properties.  Recent efforts have focused on the decoherence
 effects of a heat bath, which has also been examined in detail in the study
 of quantum brownian
 motion. Decoherence is identified as the (effective) suppression of
 interference terms in the density operator
$(\rho(x,x'),\, x\ne x')$. It has been pointed out that most
 of the models which have been considered
 are somewhat simplistic and cannot reproduce the
phenomenological features expected of a system which interacts locally
 with a homogeneous and isotropic environment\cite{GF}.  In this paper I
describe the
perceived shortcomings of existing models and illustrate the construction
of a phenomenological quantum master equation which contains many features
expected from local coupling to a homogeneous environment\cite{Gallis2}.

Although decoherence is the most interesting feature of the effects of a heat
 bath, dissipation (and
other effects) also generally appear in the dynamics of the density operator:
\begin{eqnarray}
{\partial \rho (x,x';t) \over \partial t}={\rm Hamiltonian\enskip terms}
+{\rm Dissipation\enskip terms}+\cdots-g(x,x')\rho (x,x';t).
\end{eqnarray}
The decoherence term appears as a (spatially dependent) decay
 term in the evolution equation, and
can be understood in terms of effective fluctuating forces, or potentials:
\cite{Gallis1,Diosi}
\begin{eqnarray}
g({ x},{y})  =({1\over{\hbar^2}})(c({ x};{x}) +c({ y};{y})
-2c({ x};{y}),\,
\langle V({ x},t)V({ y},s)\rangle =c({ x};{y})\delta (t-s).
\end{eqnarray}
  Typical models have a quadratic form, $g({ x},{y}) \propto (x-y)^2 $
for the decoherence term, corresponding to a fluctuating force which
 is independent of
position.  However, for a local bath one expects the correlation
 function to die off at some
characteristic length scale (the correlation length of the environment),
which has some important
ramifications for decoherence.  For a quadratic form of decoherence,
 the decay rate of the
interference terms in the density matrix increases without bound, while for
 a local model the decay
rate saturates at separations (between $x$ and $x'$) much larger than the
 correlation length of the
environment, reflecting the independence of environment fluctuations at
 large separations.  As it turns out,
 the quadratic form can be considered a short length scale
 approximation of a more detailed model.

 To consider the decoherence effects of an environment, simultaneous treatment
 of dissipation is necessary since decoherence and dissipative effects
both generally arise from the same source (the interaction with a heat bath).
  For simplicity, I  consider only linear dissipation, that is
\begin{eqnarray}
m\dot{\bf x} ={\bf p},{\hskip 1em}
\dot{\bf p} =-{\eta  \over m}{\bf p} +{\bf F}.
\end{eqnarray}

As an example of quantum dissipative evolution, Dekker\cite{Dekker} has
constructed
 a phenomenological master equation which includes
 ohmic dissipation and quadratic decoherence:
\begin{eqnarray}
{\partial \rho  \over \partial t} =& {1 \over i\hbar}[H,\rho ]
-i{\lambda  \over \hbar}[x,\{p,\rho \}]
&+{{({D}_{xp}+{D}_{px})} \over {\hbar}^{2}}[x,[p,\rho ]]
-{{D}_{xx} \over {\hbar}^{2}}[p,[p,\rho ]]-{{D}_{pp} \over
{\hbar}^{2}}[x,[x,\rho ]].
\end{eqnarray}
   The Caldeira-Leggett\cite{Caldeira} master equation is obtained
from a first-principle calculation for the effects of a simple thermal bath.
With an appropriate choice of  parameters for
 the Dekker model, the Calderia-Leggett
 master equation can be reproduced.

%%%%%%%%%%%%%%%%%%%%%%%%%%
% slide 2
%%%%%%%%%%%%%%%%%%%
Many open system models can reduce
 to the same classical phenomenology, particularly in the Markov regime,
and yet have significant differences for a quantum sytem in that same regime.
To illustrate this ``richness'' of quantum dissipative models, consider a
rather
 generic oscillator bath
model (following Zwanzig\cite{zwanzig}):
\begin{eqnarray} L={1 \over 2}m{\dot{x}}^{2}-U({\bf
x})+\sum\nolimits\limits_{\mu } {{m}_{\mu }
\over 2}[{\dot{q}}_{\mu }^{2}-{\omega }_{\mu }^{2}({q}_{\mu
}-{a}_{\mu }({\bf x}){)}^{2}].
\end{eqnarray}
 The classical calculations (the results of
 which are presumably reproduced
in at least some limit of the quantum model) are relatively straightforward.
  The classical
fluctuation-dissipation relation
  between the fluctuating forces and the
 nonlinear dissipation kernal
emerges naturally, and in the usual Markov limit becomes:
\begin{eqnarray}
\langle {f}_{i}({\bf x},t){f}_{j}({\bf y},s)\rangle
={k}_{B}T {\eta
}_{ij}({\bf x},{\bf y};t-s) =
{\overline{\eta }}_{ij}({\bf x},{\bf y})2{k}_{B}T\delta (t-s),
\end{eqnarray}
 and a simple langevin equation can
 (at least in principle)
be obtained:
\begin{eqnarray}
{\ddot{ x}}_{i}(t)=-{\partial U({\bf x}(t)) \over \partial
{x}_{i}}+{f}_{i}(
{\bf x}(t),t)-{\overline{\eta }}_{ij}({\bf
x}(t)){\dot{x}}_{j}.
\end{eqnarray}
For a homogeneous environment, the dissipation constant would be
 independent of position.

Some observations about the Markov limit are in order.  For the classical
picture,
 the spatial
correlations of the fluctuating forces are irrelevant.  After all, the particle
 can only be in one place at
one time.  For a quantum system one must consider superpositions between
 the particle at
different locations, i.e. superpositions between different trajectories for
 the particle.  My point is that
different models may produce the same classical phenomenology,
 but have some important
differences for the quantum case, in particular for the
 effective decoherence due to the
environment.

In order to help motivate some choices which will be required for the
construction of
 the new model, consider a
particle locally coupled to a scalar field.  This particular model is a natural
extension of one
considered by Unruh and Zurek\cite{UZ}. The action for this model is given by:
\begin{eqnarray}
L =&\int_{}^{}{ d}^{n}r \left\{{{1 \over 2}\left[{{\dot{ \phi }}
^{ 2}-{ c}^{ 2}({\nabla }_{ r}\phi  {)}^{2}}\right]+ \delta
  ({\bf r} -{\bf x} )\left[{{ m{\dot{\bf x}}^{2} \over 2}-
 \varepsilon \phi  ({\bf r}, t )- V ({\bf x} )}\right]}
\right\}.
\end{eqnarray}
  This model produces approximately ohmic
dissipation in one dimension\cite{Caldeira,Gallis3}. In addition, one can
extract from the
 influence functional the
effective correlation function of the fluctuating forces\cite{Gallis1,Gallis3}:
\begin{eqnarray}
 <{\bf  F}({\bf  x}, \tau  )\cdot {\bf  F}({\bf  y}, s )>0
={\hbar{ \varepsilon }^{ 2} \over 2(2 \pi { )}^{ d}}\int_{ }^{}{ d}^{d}k {k}^{
2}
\left\{{\coth({ \beta  \hbar \omega  \over  2}) \over  \omega } \cos( \omega t
)
\cos({\bf  k}\cdot({\bf  x}-{\bf y}))\right\}.
\end{eqnarray}
  This correlation function results from
independent contributions from each mode of oscillation of the field.
  With some of the characteristics suggested by this local environment in mind,
 I now turn to the actual construction of the model.

\section{The Phenomenological Model}
%%%%%%%%%%%%
%  slide 3
%%%%%%%%%%%

The initial form of the evolution of the density operator is taken to
 be in the Lindblad\cite{Lindblad} form (Schr\"odinger picture):
\begin{eqnarray} {\partial \rho  \over \partial t}=L[\rho ]
 ={1 \over i\hbar}[H,\rho ]+{1 \over
2\hbar}\sum\nolimits\limits_{\mu } [{V}_{\mu }\rho
,{V}_{\mu }^{\dagger
}]+[{V}_{\mu },\rho {V}_{\mu }^{\dagger }]
={1 \over i\hbar}[H,\rho ]+\Delta L[\rho ],
\end{eqnarray}
 for which there is a corresponding form for the Heisenberg picture
$ {L}^{{}^*}[O]$ which can readily be obtained from the cyclic
 properties of the trace.
  For a finite dimensional Hilbert space, this form is the most general
 for a completely positive dynamical
semigroup.  For infinite dimensional Hilbert spaces, it is a reasonable
 starting point.  I will be
focusing on the nonunitary part of the evolution, $\Delta L$.

The construction of the model is essentially the determination of the
operators ${V_\mu}$, subject to the
constraint that the dissipation is ohmic (expressed as an operator condition).
  This constraint
produces the ``correct'' classical phenomenology, but does not completely
 determine the model.
However, linear dissipation almost forces the ${V_\mu}$ to be at most linear in
 momentum, that is
\begin{eqnarray}
V_\mu ={A}_{\mu }({\bf x})-{\bf B}_{\mu }({\bf x})\cdot
{\bf p}.
\end{eqnarray}
Homogeneity and isotropy also serve to constrain the model.  Assuming
 some sort of mode by
mode interaction with a field, a reasonable choice is given by:
\begin{eqnarray}
\{V_\mu\} &=\{\alpha(k)e^{i{\bf k}\cdot{\bf x}} -\beta(k)e^{i{\bf k}\cdot{\bf
x}}
{\bf k}\cdot {\bf p}\}.
\end{eqnarray}
The discrete index $\mu$ has been replaced by the continuous
index ${\bf k}$.
The model is then completely specified by the complex functions $\alpha$
 and $\beta$.

The resulting nonunitary contribution to the Schr{\"o}dinger equation is given
by
 the expression:
\begin{eqnarray}
\Delta L[\rho ]&=-\int_{}{d}^{d}k{|\alpha
(k){|}^{2} \over \hbar}(\rho -{e}^{i{\bf k}\cdot {\bf x}}
 \rho {e}^{-i{\bf k}\cdot {\bf x}})
-\int_{}{d}^{d}k{|\beta (k){|}^{2} \over \hbar}({1 \over
2}\{({\bf k}\cdot {\bf p}{ )}^{2},\rho \}- e^{i{\bf k}\cdot {\bf x}}
{\bf k}\cdot {\bf p}
\rho {\bf k} \cdot {\bf p} e^{-i{\bf k}\cdot {\bf x}})\hfill\cr
&-\int_{}{d}^{d}k{{\rm Re}[\alpha (k{)}^{{}^*}\beta (k)]
\over \hbar}(e^{i{\bf k}\cdot {\bf x}}\{{\bf k}\cdot {\bf
p} , \rho  \}e^{-i{\bf k}\cdot {\bf x}})
-\int_{}{d}^{d}k{i{\rm Im}[\alpha (k{)}^{{}^*}\beta (k)]
\over \hbar}(e^{i{\bf k}\cdot {\bf x}}
[{\bf k}\cdot {\bf p}, \rho ]{e}^{-i{\bf k}\cdot {\bf x}}).\hfill
\end{eqnarray}
The position representation of the new model is given by:
\begin{eqnarray}
{ \partial  \rho  ( x , x '; t ) \over \partial  t} &=& {\rm \, Hamiltonian \,
terms}
-\left({\int_{}^{} dk{ | \alpha ( k ){ |}^{2} \over \hbar}(1- \cos  k ( x - x
'))}\right) \rho  ( x , x '; t )\cr
&-&\left({{2 \over \hbar}\int_{}^{} dk{\rm Re} [{ \alpha }^{ {}^*}( k ) \beta (
k ) ] k  \sin  k ( x - x ')}\right)\left({{\partial  \over \partial  x}
-{\partial  \over \partial  x '}}\right) \rho  ( x , x '; t )\cr
&-&\left({ i\int_{}^{}dk{\rm Im} [{ \alpha }^{ {}^*}( k ) \beta  ( k )] k  \sin
k ( x - x ')}\right)\left({{\partial  \over \partial  x} +{\partial  \over
\partial  x '}}\right) \rho  ( x , x '; t )\cr
&+&\left({\int_{}^{} dk | \beta ( k ) { |}^{2}\hbar{ k}^{
2}}\right)\left({{{\partial }^{2} \over \partial {x}^{2}}-{{\partial }^{2}
\over
\partial {x'}^{2}}}\right) \rho  ( x , x '; t )\cr
&+&\left({\int_{}^{} dk | \beta ( k ){ |}^{2}\hbar{ k}^{ 2}\cos
k(x-x')}\right)\left({{{\partial }^{2} \over \partial x\partial x'}}\right)
\rho
( x , x '; t ).
\end{eqnarray}
The first nonhamiltonian term is responsible for decoherence.  The
corresponding
 noise spatial correlation is determined by $\alpha (k)$.  The characteristic
 length should be on the order of the inverse of the ``width'' of
$|\alpha (k)|^2$ in $k$ space.  The second nonhamiltonian term
is responsible for the dissipation.  Clearly the
dissipation and other terms are more complicated in this new model. However,
 that would also be expected from a more elaborate first principle
 calculation.

%%%%%%%%%%
% slide 4
%%%%%%%%
By examining the Eherenfest relations of physical quantities using $
{L}^{{}^*}$
, some interesting
 physical features of the new model emerge.  By construction, the average
 position and momentum obey relations corresponding to ohmic dissipation:
\begin{eqnarray}
 {d\langle P\rangle \over dt}=
{i \over {\hbar}}\langle[H,P] \rangle -{\eta \over m}\langle P\rangle,
{\hskip 1em}
{d\langle x\rangle \over dt}=
{i \over {\hbar}}\langle[H,x] \rangle ={\langle P\rangle\over m},
\end{eqnarray}
where
\begin{eqnarray}
 {\eta  \over m}=\int_{}dk 2{\rm Re}({\alpha}^{{}^*}(k)\beta
(k))k^{2}.
\end{eqnarray}
With only limited constraints on $\alpha$ and $\beta$ ($\gamma$
 must be positive), the kinetic
 energy is seen to be thermalized:
\begin{eqnarray}
{d\over dt}\langle ({{P}^{2} \over 2m}-{{k}_{B}T \over 2}) \rangle =
\langle{i \over {\hbar}}[H,({{P}^{2} \over 2m}-{{k}_{B}T \over 2})] \rangle
 -\gamma \langle ({{P}^{2} \over 2m}-{{k}_{B}T \over 2})\rangle,
\end{eqnarray}
where
\begin{eqnarray}
\gamma \equiv  2{\eta  \over m}-
\int_{}{d}^{d}k{|\beta
(k){|}^{2}
\over d}\hbar{k}^{4}, \,\,
{{k}_{B}T \over 2} \equiv  {1 \over \gamma
}\int_{}{d}^{d}k{|\alpha
(k){|}^{2}
\over 2m}\hbar{k}^{2}.
\end{eqnarray}
  The effective temperature is determined by $\alpha$ and $\beta$.

%%%%%%%%%%%%%
% Slide 5
%%%%%%%%%%%

A low length scale approximation of the new model can be obtained by
 expanding the exponential terms in powers of ${\bf k} \cdot {\bf x}$:
\begin{eqnarray}
\Delta L[\rho ]&\cong& -\int_{}{d}^{d}k{|\alpha (k){|}^{2}
\over 2\hbar}[{\bf k}\cdot {\bf x},[{\bf k}\cdot {\bf x},\rho]]
-\int_{}{d}^{d}k{|\beta (k){|}^{2} \over 2\hbar}[{\bf
k}\cdot {\bf p} ,[{\bf k}\cdot {\bf p} ,\rho ]]\hfill\cr
&-&\int_{}{d}^{d}k{i{\rm Re}(\alpha (k{)}^{{}^*}\beta (k))
\over \hbar}[{\bf k}\cdot {\bf x}
 ,\{{\bf k}\cdot {\bf p} ,\rho \}]
+\int_{}{d}^{d}k{{\rm Im}(\alpha (k{)}^{{}^*}\beta (k))
\over \hbar}[{\bf k}\cdot {\bf x} ,[{\bf k}\cdot {\bf p} ,\rho ]].\hfill
\end{eqnarray}
  The lowest nonvanishing terms are second order, which exactly reproduces
the Dekker master equation for 1 dimension.  As a result, we can think of
 the Dekker or Caldeira-Leggett equations as a low length scale approximation
 for more general models.

On the other hand, the Caldeira-Leggett master equation,
\begin{eqnarray}
 \Delta L[\rho ]={\eta  \over i2m\hbar}[x,\{p,\rho \}]-{\eta
{k}_{B}T
\over
{\hbar}^{2}}[x,[x,\rho ]],
\end{eqnarray}
 can be considered a special case of the
 Dekker master equation, with the $D_{xp}$ terms equal to zero (which Dekker
has
 argued should be the case) and an additional low momentum approximation which
 ignores the $D_{xx}$ term.  With this type of special case in mind, we can
 construct a low momentum approximation for the new model which includes only
 the decoherence and dissipation terms:
\begin{eqnarray}
\Delta L[\rho ]=&-\int_{}{d}^{d}k{|\alpha (k){|}^{2} \over
\hbar}(\rho
-{e}^{i{\bf k}\cdot {\bf x}} \rho {e}^{-i{\bf k}\cdot {\bf
x}}
)
&-\int_{}{d}^{d}k{{\rm Re}(\alpha (k{)}^{{}^*}\beta (k))
\over
\hbar}(\{{\bf
k}\cdot {\bf
p},{e}^{i{\bf k}\cdot {\bf x}} \rho {e}^{-i{\bf k}\cdot {\bf
x}}\}).
\end{eqnarray}
 This would seem to be a
 likely starting point for applications of this model.  However, this
 approximation is not a positive form for the dynamics.
%%%%%%%%%%%
% slide 6
%%%%%%%%%%%%%%

Finally, I would like to look at the Wigner representation of the new model,
 which has some interesting features.  If we expand the terms of the evolution
 equation in powers of $\hbar$ (in the same manner as is typically done with
the
 potential),
\begin{eqnarray}
{\dot W}(q,p)=&
-&{1 \over  m}{ \partial  \over \partial q}(pW)+{\partial  \over \partial
p}(V'(q)W)
 +\sum\nolimits\limits_{n=1}^{\infty } {(\hbar{)}^{2n}(-1{)}^{n}{2}^{-2n} \over
(2n)!}{V}^{(2n+1)}{{\partial }^{2n+1} \over \partial {p}^{2n+1}}W\cr
&+&\lambda { \partial  \over \partial p}(pW)
+\sum\nolimits\limits_{n=1}^{\infty } (\hbar{)}^{2n}\left({\int_{}^{}dk{2{\rm
Re}({\alpha }^{{}^*}\beta ){k}^{2n+1} \over (2n+1)!}}\right){{\partial }^{2n+1}
\over \partial {p}^{2n+1}}W\cr
&+&{ D}_{pp}{{ \partial }^{2} \over \partial {p}^{2}}W
+\sum\nolimits\limits_{n=1}^{\infty }
(\hbar{)}^{2n+1}\left({\int_{}^{}dk{|\alpha {|}^{2}{k}^{2n+2} \over
(2n+2)!}}\right){{\partial }^{2n+2} \over \partial {p}^{2n+2}}W\cr
&+&({ D}_{xp} +{ D}_{px} ){{\partial }^{2} \over \partial p\partial q}W
+\sum\nolimits\limits_{n=1}^{\infty } (\hbar{)}^{2n+1}\left({\int_{}^{}dk{{\rm
Im}({\alpha }^{{}^*}\beta ){k}^{2n+2} \over (2n+1)!}}\right){\partial  \over
\partial q}{{\partial }^{2n+1} \over \partial {p}^{2n+1}}W\hfill\cr
&+&{ D}_{xx}{{ \partial }^{2} \over \partial {q}^{2}}W
+\sum\nolimits\limits_{n=1}^{\infty }
(\hbar{)}^{2n-1}\left({\int_{}^{}dk{|\beta
{|}^{2}{k}^{2n+2} \over (2n)!}}\right){{\partial }^{n} \over \partial
{p}^{n}}\left({{{\hbar}^{2} \over 4}{{\partial }^{2} \over \partial
{q}^{2}}+{p}^{2}}\right)W,
\end{eqnarray}
 the lowest order terms correspond exactly to the Wigner representation
 of the Dekker equation.  The Wigner representation of the Dekker equation is
 a standard classical type of diffusion equation.  This illustrates the idea
 that the ``classical'' nature of the system emerges when coherent
superpositions
 are not important in the dynamics.  In this case, the relevant superpositions
 are between different locations separated by distances on the order of the
 environment correlation length.

The convolution theorem can also be used to write down the Wigner
 representation of the evolution:
\begin{eqnarray}
{\dot W}(q,p)&=&{\rm (Hamiltonian \enskip terms)}
-{2 \over \hbar}\int_{}^{}dp'(p-p')W(q,p-p'){p' {\rm Re}[{\alpha }^{{}^*}\left(
{p' \over \hbar}\right) \beta\left({p' \over \hbar}\right) ]\over
{\hbar}^{2}}\hfill\cr
&-&\left({\int_{}^{}dk{|\alpha {|}^{2}(k) \over
\hbar}}\right)W(q,p)+\int_{}^{}dp'W(q,p-p'){|\alpha \left({p' \over
\hbar}\right) {|}^{2} \over {\hbar}^{2}}
%\smallskip
\cr
&-&{\partial  \over \partial q}\int_{}^{}dp'W(q,p-p'){p' {\rm Im}[{\alpha
}^{{}^*}\left({p' \over \hbar}\right)\beta\left({p' \over \hbar}\right) ]\over
{\hbar}^{2}}\hfill\cr
&+&{D}_{xx}({1 \over 4}{{\partial }^{2} \over \partial {q}^{2}}-{{p}^{2} \over
{\hbar}^{2}})W(q,p)
+\int_{}^{}dp'{{p}^{2} |\beta\left({{p' \over \hbar}}\right) {|}^{2}\over
{\hbar}^{2}}({1 \over 4}{{\partial }^{2} \over \partial {q}^{2}}+{{(p-p')}^{2}
\over {\hbar}^{2}})W(q,p-p').\hfill
\end{eqnarray}
  One apparent effect in the new model
 is a spreading induced by these convolution terms.

In summary, a new phenomenological master equation for ohmic dissipation
and decoherence has been constructed which has completely positive dynamics.
The new model has the
features expected from local coupling to a homogeneous environment:
specifically,
the evolution is istropic and translationally invariant.  Spatial correlations
of the
environment appear explicitly in the models.  The new model also includes
previous
results as low length scale approximations.

\end{document}